# Segment anything model for head and neck tumor segmentation with CT, PET and MRI multi-modality images

Jintao Ren[1,2], Mathis Rasmussen[1,2], Jasper Nijkamp[1,2], Jesper Grau Eriksen[1,3] and Stine Korreman[1,2]

[1]Department of Clinical Medicine, Aarhus University, Aarhus, Denmark
[2]Danish Center for Particle Therapy, Aarhus University Hospital, Aarhus, Denmark
[3]Department of Experimental Clinical Oncology, Aarhus University Hospital, Aarhus, Denmark

**Abstract** Deep learning presents novel opportunities for the auto-segmentation of gross tumor volume (GTV) in head and neck cancer (HNC), yet fully automatic methods usually necessitate significant manual refinement. This study investigates the Segment Anything Model (SAM), recognized for requiring minimal human prompting and its zero-shot generalization ability across natural images. We specifically examine MedSAM, a version of SAM fine-tuned with large-scale public medical images. Despite its progress, the integration of multi-modality images (CT, PET, MRI) for effective GTV delineation remains a challenge. Focusing on SAM's application in HNC GTV segmentation, we assess its performance in both zero-shot and fine-tuned scenarios using single (CT-only) and fused multi-modality images. Our study demonstrates that fine-tuning SAM significantly enhances its segmentation accuracy, building upon the already effective zero-shot results achieved with bounding box prompts. These findings open a promising avenue for semi-automatic HNC GTV segmentation.

## 1 Introduction

In radiotherapy, delineating the gross tumor volume (GTV) for head and neck cancer (HNC) is a critical yet challenging component of treatment planning. This complexity arises due to the proximity of adjacent organs at risk (OARs) within a constrained space and the tumors' heterogeneous appearance. Manual GTV delineation is not only time-consuming and dependent on a high level of expertise but also requires the integration of multi-modality images such as CT, PET, and MRI. Furthermore, the process is prone to inter-observer variability, which can significantly impact treatment outcomes. Consequently, there is a pressing need for a fully or semi-automatic approach that can streamline the delineation process, enhance consistency, and improve clinical workflow.

Deep learning segmentation models are increasingly recognized for their ability to discern complex image features from extensive datasets of historical images and annotations. Despite their promise, the application of HNC GTV segmentation often necessitates significant manual adjustments for certain patients[1]. Simplifying the definition of a region of interest (ROI) through bounding boxes has been shown to markedly improve outcomes[2, 3]. Furthermore, incorporating 'human-in-the-loop' interactions not only enhances model accuracy but also its interpretability[4, 5, 6].

Nonetheless, a major hurdle for these models is their struggle with generalization, particularly when there is a significant divergence between training and testing datasets. These could be for example trained on oropharyngeal tumors and tested on larynx tumors, or inter-modality inferences. The Segment Anything Model (SAM) has emerged as a promising solution, demonstrating exceptional zero-shot learning capabilities, trained on over 1.1 billion segmentation masks and roughly 11 million natural images[7]. SAM supports user prompts via points, bounding boxes, and text, yet its successful deployment in medical imaging (e.g., X-ray, CT, MRI) is hampered by the substantial domain gap between natural and medical images[8].

To bridge the domain gap, recent efforts have focused on fine-tuning both public and private medical image datasets[9, 10]. A notable contribution by Ma et al.[11] introduced MedSAM, a version of the SAM fine-tuned with over 1.5 million medical image-mask pairs, demonstrating significant promise in HNC tumor segmentation using CT images. However, GTV delineation in most hospitals involves integrating multiple registered image modalities (e.g., CT, PET, MRI). Given that SAM was initially trained on natural images with red-green-blue (RGB) channels, and MedSAM was fine-tuned for single-modality scenarios, the challenge of fusing registered multi-modality images remains. Consequently, there may be reservations regarding the efficacy of MedSAM's zero-shot inference on local multi-modality datasets.

In this study, we explore MedSAM's capabilities in segmenting both primary tumors (GTV-T) and the involved lymph nodes (GTV-N) across single- and multi-modality images. We assess the model's performance in zero-shot and fine-tuned scenarios on our local dataset.

## 2 Materials and Methods

A total of 576 HNC patients from Aarhus University Hospital treated with radiotherapy in the period 2013-2020 were included in this study. Each patient had undergone treatment planning with FDG-PET/CT, T1-weighted (T1w), and T2-weighted (T2w) MRI. The T1w and T2w images were initially registered to the CT images using rigid and then deformable registration methods. The dataset was randomly divided into three subsets: 470 for training, 94 for validation, and 97 for testing. The clinical treatment planning process involved referencing all image modalities to delineate both GTV-T and GTV-N, with these clinical delineations serving as the ground truth.

*2.1 Single- and Multi-modality*
Although MedSAM is trained on datasets from various modalities, like the original SAM, it processes inputs from only a single modality at a time. For inputs involving a single modality, we adopted the approach of replicating the CT images across three channels to mimic RGB inputs, a method we refer to as 'CT-only.'

For multi-modality fusion, to accommodate four image modalities while maintaining SAM's simplicity, we devised a unique strategy: CT images were assigned to the red channel, PET images to the green channel and a blend of T1w and T2w MRI images to the blue channel. To integrate the T1w and T2w images, we first normalized them to the same intensity range and then computed the average intensity for each corresponding pixel pair. This approach is termed 'multimodal'.

*2.2 Preprocessing and bounding box prompt*
For preprocessing, CT images were adjusted with a window level of 40 and a window width of 400 to highlight soft tissue. PET images were clipped to a range of 0-6 SUV[12], while MRI images were adjusted to include intensities from the 0th to the 99.5th percentile of their maximum values. Subsequently, all images from these different modalities were rescaled to a 0-255 range, aligning them with the typical intensity range of natural images. After combining these into RGB channels, each 2D slice was normalized to float values between 0 and 1, preparing them as input for SAM.

Regarding the segmentation label masks, we treated each GTV-T and GTV-N object within the 2D slices as distinct segmentation targets. For SAM's prompting mechanism, we utilized a bounding box strategy. On each 2D slice, we introduced a random perturbation margin ranging from 0 to 20 mm to box out each GTV instance.

*2.3 Zero-shot and finetuning of MedSAM model*
We utilized the lite version of MedSAM, which employs a tiny vision transformer (TinyViT) model distilled from the larger variant, as our pre-trained SAM model base[13]. To evaluate its zero-shot generalization ability on our local dataset, we performed inference on our test set without any additional training.

For fine-tuning, the pre-trained lite MedSAM model was further trained on our local dataset for up to 20 epochs. We saved the epochs of the model that achieved the lowest Dice Loss on the validation set to ensure optimal performance.

We followed MedSAM's training protocol, employing a hybrid loss function that combines Dice loss, cross-entropy, and IoU loss. The AdamW optimizer was used, with a batch size of 8 and an initial learning rate of 3e-4. This rate was scheduled to reduce by a factor of 0.9 upon hitting a performance plateau, with patience of 5 epochs. The training was executed on two NVIDIA A40 GPUs, each equipped with 40 GB of GPU memory.

For the comparative analysis, we utilized two fully automated configurations of nnUNet[14], processing all image modalities (four image modalities) with the identical data split as SAM finetune. The first configuration functioned without a bounding box (nnUNet w/o box), whereas the second configuration included a bounding box as an additional input channel for each 2D slice, aligning with the approach taken by MedSAM (nnUNet w/ box). Both nnUNet configurations were trained for 1000 epochs in 3D full resolution.

*2.4 Evaluation*
To evaluate segmentation performance, we computed the mean and one standard deviation for the Dice Similarity Coefficient (DSC) as the primary metric, complemented by the 95% Hausdorff Distance (HD95). These evaluations were conducted on the entire 3D volumes of the images on the test set, comparing the segmented outputs with the ground truth for both GTV-T and GTV-N. Additionally, we monitored the fine-tuning time required for each epoch and the inference time needed to predict each slice.

## 3 Results
In assessing segmentation performance, we evaluated MedSAM under both zero-shot and fine-tuned conditions, using either CT-only or fused multimodal inputs. We also compared these results against the fully automatic multi-modality nnUNet, both in its standard configuration and an enhanced setup incorporating a bounding box. The outcomes, including both DSC and HD95 metrics, are detailed in Figure 1.

When examining zero-shot outcomes, multimodal inputs slightly outperform CT-only for GTV-T (0.70 vs. 0.69), whereas CT-only surpasses multimodal for GTV-N (0.75 vs. 0.68). However, following fine-tuning, both CT-only and multimodal approaches see a significant boost in segmentation accuracy, as reflected in DSC scores. Specifically, for CT-only inputs, DSC scores increased from 0.69 to 0.82 for GTV-T and from 0.75 to 0.85 for GTV-N. Similarly, multimodal inputs saw DSC improvements from 0.7 to 0.82 for GTV-T and from 0.68 to 0.85 for GTV-N with fine-tuning.

A nnUNet model without a bounding box exhibits lower performance (0.68 for GTV-T and 0.63 for GTV-N) and higher variance, however, the inclusion of bounding boxes of each target at each 2D slice in nnUNet markedly improves results (0.88 for GTV-T and 0.89 for GTV-N). The HD95 exhibits a trend similar to that observed with DSC.

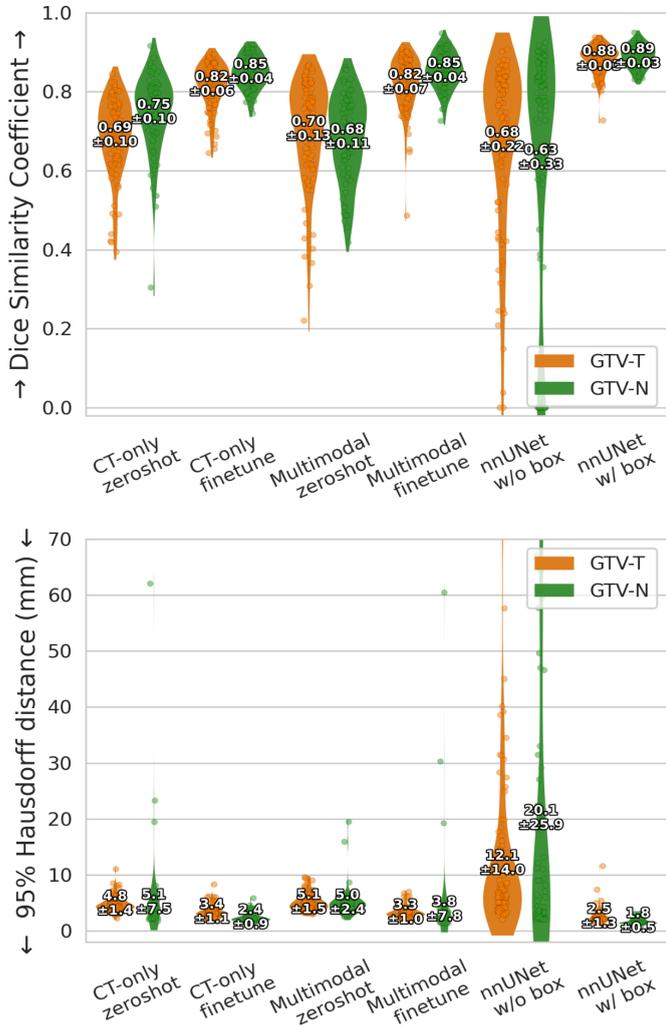

Figure 1: Comparative analysis of segmentation performance across groups for GTV-T and GTV-N, illustrated through violin plots using the Dice Similarity Coefficient (DSC) and the 95% Hausdorff Distance (HD95).

Figure 2: A case study illustration of CT-only and multimodal approaches on zero-shot and fine-tuned MedSAM for GTV-T and GTV-N segmentation.

The fine-tuning time for a lite MedSAM model using the training set (470 patients) averaged 11 minutes per epoch. We obtained the optimal fine-tuned models at epoch 8 for CT-only inputs and at epoch 4 for multimodal inputs. The prediction time for a target, given a bounding box on a 2D slice, was approximately 0.03 seconds, with no discernible difference between CT-only and multimodal inputs.

## 4 Discussion

In this investigation, we systematically explored the efficacy of the SAM model in segmenting GTV-T and GTV-N using local single- and multi-modality images. Even though the ground truth delineation referenced multimodality images, the CT-only input yielded comparable results for zero-shot inference with MedSAM on GTV-T. Interestingly, it achieved better results on GTV-N. The finetuning shows the effectiveness of enhancing segmentation accuracy and consistency for both groups.

Previous studies have demonstrated that fully automatic U-Net configurations, which combine multiple modalities (e.g., CT and PET), significantly outperform those using a single modality[15]. In our study, the CT-only approach with MedSAM achieved better segmentation accuracy compared to the fully automatic multi-modality nnUNet. This suggests that the inclusion of a bounding box for prompting effectively localizes the region of interest, even in zero-shot inference scenarios, as evidenced by the significant reduction in outliers highlighted by the HD95 metric. This benefit is also apparent in nnUNet's performance when utilizing bounding boxes. Conversely, the similar overall performance between CT-only and multimodal approaches in zero-shot inference could imply that the method used for fusing multimodal images may not be fully effective by SAM. This aligns with prior research emphasizing PET's crucial role in directing the network's focus to the relevant area[4,12]. Additionally, there are instances, where multimodal could enhance GTV-T boundary contrast, as depicted in Figure 2.

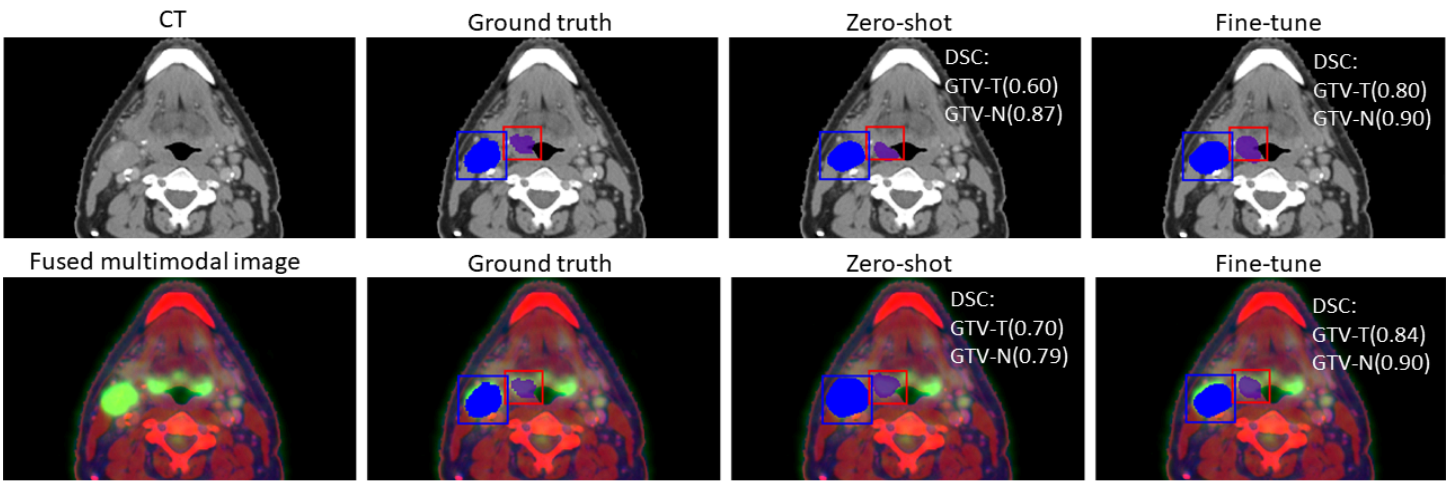

The significant improvement in performance following fine-tuning demonstrates the strong adaptation capability of SAM to new datasets within just a few epochs of training (8 for CT-only and 4 for multimodal). This is particularly notable in our proposed multimodal fusion strategy, considering that neither SAM nor MedSAM had previously encountered such image types. Despite their robust generalization capabilities, a gap remains between the fine-tuned models and the nnUNet when using the same bounding boxes. This gap may be attributed to nnUNet's powerful feature learning capability, stemming from its large model size.

These findings also spotlight a key limitation of fully automatic segmentation methods: their diminished focus when predicting GTV with full 3D image size. Interactions with models like nnUNet, while robust, can be more complex and time-intensive compared to simpler approaches like SAM. Predicting a comprehensive 3D head and neck image incorporating CT, PET, T1w, T2w, and an additional channel for human inputs can take up to 2 minutes when using nnUNet. In contrast, using a bounding box to prompt SAM for a single slice yielded instant responses (~0.03 seconds). Therefore, integrating SAM with U-Nets as a semi-automated annotation tool holds great promise for incorporating 'human-in-the-loop' practices in clinical settings. Such tools could not only increase the credibility and interpretability of AI models but also enable personalized treatment planning for optimized radiotherapy[4].

Our study has some limitations. Firstly, we contrasted the 2D MedSAM model with the 3D nnUNet, where the latter might benefit from contextual dependencies along the axial direction. However, given that SAM is designed for interactive use and nnUNet for fully automatic use, a direct comparison may not constitute a fair contest. The nnUNet, when equipped with a bounding box for each slice, sets a high benchmark for semi-automatic segmentation, whereas its performance without a bounding box represents a decent threshold for fully automatic segmentation. Secondly, while we selected the best-performing fine-tuned model based on its performance on the validation set to minimize the risk of overfitting, there remains a potential for catastrophic forgetting when applied to other datasets. Validating the model on a public HNC dataset could help mitigate this concern. Moreover, future studies exploring the integration of various prompting methods, including clicks, brushes (masks), or textual (language) inputs, in addition to bounding boxes, would be valuable.

## 5 Conclusion

In conclusion, our study demonstrated the potential of zero-shot and fine-tuning SAM for the segmentation of GTV in head and neck cancer. We found that fine-tuning significantly enhances segmentation accuracy for both CT-only and fused multimodal images, underscoring the model's robust adaptation capabilities. Our findings suggest promising avenues for enhancing semi-automatic segmentation tools in clinical settings using SAM.